# Water Aging Effects on Graphene Nanoplatelets and Multi-walled Carbon Nanotube Reinforced Epoxy Glass Fiber Nanocomposites

B. M. Madhu[1]*, T. V. Rashmi[2]

[1]*Department of EEE, VTU Research Center, Siddaganga Institute of Technology, Tumkur, Karnataka, India, [2]Department of EEE, Siddaganga Institute of Technology, Tumkur, Karnataka, India*

**ABSTRACT**

Nanocomposites reinforced with hybrid fillers of multi-walled carbon nanotubes (MWCNTs) and graphene nanoplatelets (GNPs) were developed, aimed at improving electrical and morphological properties of the hybrid nanocomposites while reducing the cost of the final product. GNPs and MWCNTs nanofillers have shown improved electrical and morphological properties with most polymers. In this work, the effect of short-term water aging for 1440 h on MCWNTs and GNPs reinforced epoxy glass fiber nanocomposites was studied. Epoxy nanocomposites were prepared with varying combinations of MCWNTs and GNPs (2:1, 2:2, and 2:3wt. %) as conducting fillers and their electrical conductivity was evaluated after short-term water aging. It was shown that the addition of MCWNTs and GNPs enhanced the electrical conductivity of composites: A low percolation threshold was achieved with 2 wt. % MCWNTs and 3 wt. % GNPs. The addition of MCWNTs enhanced the electrical conductivity and dielectric constant, confirming the synergistic effect of CNTs as multifunctional filler. Microstructural investigations and morphology of nanocomposites were investigated using Fourier transform-infrared spectroscopy and X-ray diffraction. The novelty of this work arises from the combination of two conducting fillers with different geometry and aspect ratios as well as different dispersion characteristics. The results obtained from the electrical measurements after water aging on hybrid nanocomposites indicates slight increases in electrical conductivity.

## 1. INTRODUCTION

Glass fibers/carbon fibers reinforced epoxy matrix is the materials suited for core of high-temperature low-sag conductors, which provides required functional thermal, mechanical, and electrical properties [1-3]. Epoxy resin is the good matrix for glass fiber-reinforced epoxy nanocomposites (GFRP) for various engineering applications due to their low density, good thermal, physical, electrical, and mechanical properties, besides their good processability. Hand layup technique is good for laboratory-scale production of hybrid nanocomposites, where the pultrusion technique is suitable for achieving higher mechanical and thermal properties. The excellent electrical conductivity and mechanical properties of multi-walled carbon nanotubes (MWCNTs) make them ideally suited as nanofiller for conducting polymer composites, capable of dissipating electrostatic charges, and shielding devices from electromagnetic radiation [4]. The modification of epoxy is essential as several applications which demand higher mechanical and thermal properties as well as stability under harsh conditions [5-7]. Integration of many nanosized phases into epoxy-based materials is a unique way to solve the problem. These materials added with filler materials of varying dimensions from micro to nano in the form of powder or fibers, help in improvement of specific properties. Various nanofillers, namely, carbon nanotubes, nanoclay, carbon black, and graphene nanoplatelets (GNPs) were used to enhance the properties of epoxy resins. Several research works were carried out to synthesize and fabricate MCWNTs/epoxy nanocomposites but restricted by high cost which limits applications. It found that the cost-effective GNPs could be produced using graphite flakes by exfoliating. The large surface area of GNPs and the high aspect ratio of MWCNTs allow the formation of electrical conductive network at lower filler content. The large surface area of graphene increases the contact area with the polymer matrix and maximizing the stress transfer from the polymer to the GNPs [8,9]. The components for various applications made from epoxy resins may be exposed to humidity and elevated temperatures. The mechanical and other properties of epoxy systems can be lowered due to moisture absorption. With increased temperatures (elevated temperature), the absorbed water generally works as plasticizer and thus decreases the mechanical performances. Besides, biodegradation of polymer materials, for example, by fungal growth, is likely to be increased with absorbed moisture in the materials even at ambient temperatures if porosity exists in the composite materials. Therefore, there is a need for understanding the changes in epoxy resins due to hydrothermal aging [10-12]. In this study, hydrothermal aging is conducted on GFRP epoxy nanocomposites (grapheneous nanocomposites [GNCs]). The study includes the synergetic effects of different wt. % of GNPs-MWCNTs hybrids on the properties of hybrid nanofiller coordination.

*Corresponding author:
E-mail: madhubm@sit.ac.in

This combination of GNPs-MWCNTs demonstrated synergistic improvement of mechanical properties, electrical conductivity. Fourier transform infrared (FTIR) analyses were performed to study the changes of molecular conformations in matrix specimens due to hydrothermal aging. Morphological analysis confirmed the uniform dispersion of both GNPs and MWCNTs within the epoxy matrix. For the hybrid GNPs-MWCNTs filler system (i.e., with the 2 wt. %:3 wt. % ratio), MWCNTs were seen to align themselves on the GNPs surfaces creating an interconnected strong nanofiller network in the epoxy.

## 2. MATERIALS AND METHODS

### 2.1. Materials and Fabrication

The GNPs and MWCNTs are the nanofillers and aluminum trihydrate is the microfiller used to improve the functional properties of the GNC. The matrix material is Araldite (MY 740) epoxy along with hardener Aradur (HY 918). The epoxy matrix is reinforced with E-glass fiber to fabricate using pultrusion process with pulling rate of 5–6 m/h as it supports complete wetting and uniform dispersion/growth of the nanofillers on surface of the glass fiber. Initially, nano- and microfillers were mixed with epoxy and hardener using mechanical mixer to obtain a well-dispersed composite. Then, it is transferred to temperature-controlled container in the pultrusion setup. The fiberglass is pulled through a rectangular die of the 3.7 mm void with a 40 mm width, while temperature maintained at 140°C–150°C for curing further sections of 1 m length were cut and transferred to post-curing chamber. Identification of GNC samples with composition is tabulated in Table 1.

### 2.2. Measurements

Pultruded hybrid nanocomposites were machined into rectangular samples of 10 mm side, with thickness of 3.7 mm. Samples were oven-dried for 24 h at 50°C and weighed in a high precision balance with 0.001 mg resolution; the dry samples were kept in desiccator. Weight and density of the samples were measured according to ISO 1183-1:2012 plastics – methods for determining the density of non-cellular plastics using Mettler Toledo density kit ME-DNY-43 immersion method. The crystalline phases of the prepared samples were studied by X-ray diffraction (XRD) using Philips X'pert Pro X-ray diffractometer (PW3040) with Cu-Kα radiation (λ = 1.5405 Å). The XRD data were collected at slow scan (2°/min) with a step size of 0.02° in a wide range of Bragg angles 5°C–90°C at room temperature. FTIR characterization is done using Bruker alpha Eco-ATRIR with 18 scans. Dielectric properties were measured using high precession LCR meter (HIOKI-IM3536) and (HIOKIL2000) probe with aid of precision temperature-controlled chamber in the frequency range from 10 Hz to 8 MHz with a drive voltage of 1 V for all samples. The electrical conductivity ($\sigma_{ac}$) was recorded directly.

## 3. RESULTS AND DISCUSSION

### 3.1. Morphological Characterization

The XRD peaks of GFRP epoxy nanocomposites are shown in Figure 1. From the graph, it can be observed that the peaks appeared in pattern indicate the peaks for GNPs, MWCNTs and neat epoxy glass fiber matrix are in agreement with the literature data. GNP has a major peak at 26.45° and a small broad peak at 54.61°. MWCNT has a prominent peak at 2θ value of 25.48° and a small peak at 43.13° [1].

From the XRD patterns in Figure 1, it can be observed the respective peaks of GNP and MWCNT nanofillers and in the GNC samples, these peaks are less prominent showing the homogeneous dispersion.

### 3.2. Fourier Infrared Spectroscopy Analysis

FTIR spectroscopy measurement was performed by taking powder form of the samples and was scanned in the range of 500 $cm^{-1}$–4000 $cm^{-1}$ at a resolution of 2.0$cm^{-1}$ at room temperature. It can be observed from Figure 2 that there is a change in intensity of transmittance for GFRP epoxy nanocomposites before and after water aging.

The results obtained indicate that, with aging, there is a decrease in absorption spectrum. However, as compared to neat GFRP epoxy composites, the rate of reduction was found to be less. Observations from the spectrum show a broad peak due to the presence of O–H in the region of 3400 $cm^{-1}$. The broad peak sets at around 2930 $cm^{-1}$ might be attributed to C–H stretching. A weak peak at 1730 $cm^{-1}$ indicates the presence of ester groups (carboxyl acid groups). Peak at 1090 $cm^{-1}$ is due to O-H bending [13,14].

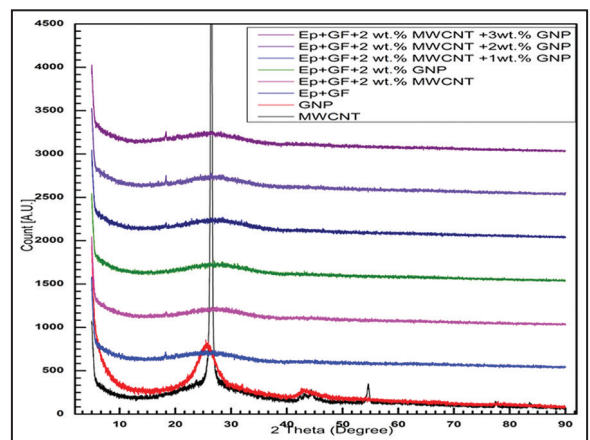

**Figure 1:** X-ray diffraction patterns of pure graphene nanoplatelets and multi-walled carbon nanotubes and grapheneous nanocomposite samples.

**Table 1:** Sample identification and composition.

| S. No. | Sample identification | Composition | Epoxy+Fillers | | | | Glass fiber |
|---|---|---|---|---|---|---|---|
| | | | MWCNTs (wt. % of Ep) | GNPs (wt. % of Ep) | ATH (wt. % of Ep) | Epoxy wt. % | |
| 1. | Ep+Gf | Epoxy/Glass fiber/ATH | -- | -- | 20 | 80 | 80 wt. % total weight of composites |
| 2. | CNT2 | Epoxy/Glass fiber/ATH/MWCNTs | 2 | -- | 20 | 78 | |
| 3. | GNP1 | Epoxy/Glass fiber/ATH/GNPs | -- | 1 | 20 | 79 | |
| 4. | H1 | Epoxy/Glass fiber/ATH/GNPs/MWCNTs | 2 | 1 | 20 | 77 | |
| 5. | H2 | Epoxy/Glass fiber/ATH/GNPs/MWCNTs | 2 | 2 | 20 | 76 | |
| 6. | H3 | Epoxy/Glass fiber/ATH/GNPs/MWCNTs | 2 | 3 | 20 | 75 | |

ATH: Aluminum trihydrate, GNPs: Graphene nanoplatelets, MWCNTs: Multi-walled carbon nanotubes

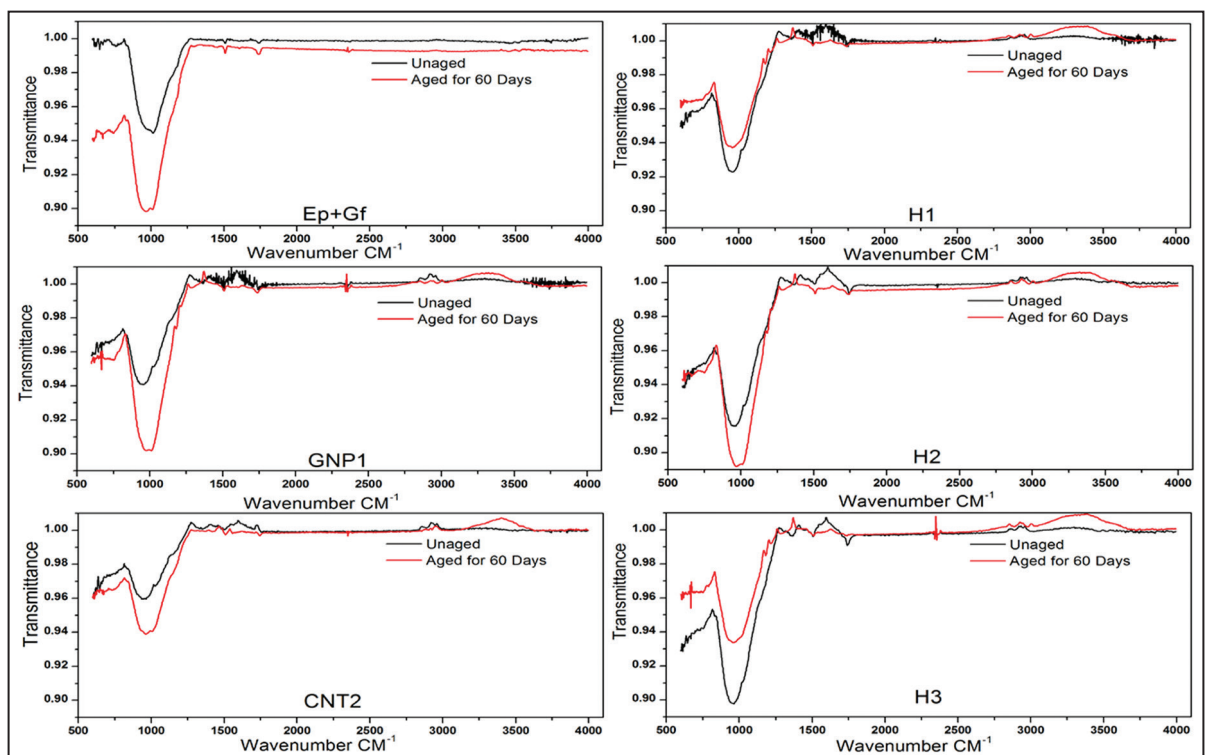

**Figure 2:** Fourier-transform infrared spectroscopy of grapheneous nanocomposites samples comparing filler loading and water aging.

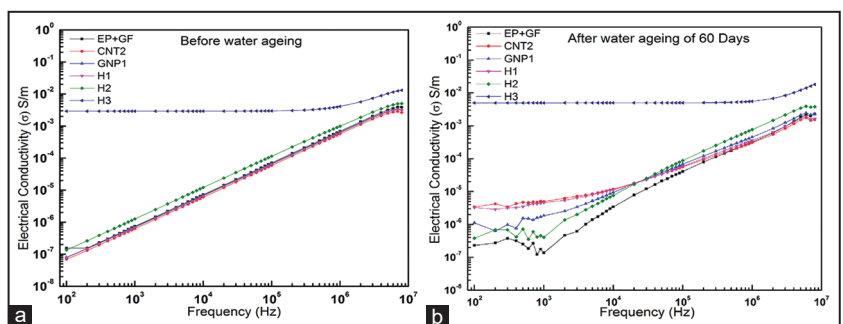

**Figure 3:** (a and b) Effects of frequency, filler loading, and water aging on AC conductivity ($\sigma_{ac}$).

### 3.3. Effects of Frequency, Filler Loading, and Water Aging on AC Conductivity ($\sigma ac$)

The dependence of AC conductivity ($\sigma_{ac}$) of composites with frequency is shown in Figure 3a and b, respectively. The effects of AC conductivity have been studied above 100 Hz-8 MHz. From Figure 3, it was observed that with increasing frequency, the $\sigma_{ac}$ of epoxy and epoxy nanocomposite increases exponentially except in H3 sample. Further, in the case of H3 sample, the increase in the conductivity is not significant. The increase in $\sigma_{ac}$ with frequency may be attributed to the hopping conduction mechanism and having a bandgap similar to that of a semiconductor. Due to localization of charge carriers in the bulk of material, formation of polarons takes place and the hopping conduction may occur between the nearest neighboring sites increasing the $\sigma_{ac}$ drastically at higher frequencies.

The GFRP epoxy nanocomposites have hydrophilic –OH bonds which attract impurities such as water, ions, or substances that are easily dissociated into ions would be introduced into the resin. From Figure 3, there is no significant change in $\sigma_{ac}$ water aging of GFRP epoxy nanocomposites which is observed. With increase in filler concentration, the hydrophobic surface of nanofillers increases more and more, confirming the reduction of the water sorption in epoxy materials. The effect of water aging on GFRP epoxy nanocomposites shows a slight increase in $\sigma_{ac}$ value. The concentration of water in the matrix may be sufficient for conduction and is likely to provide a channel for charge carriers resulting in increased conductivity. Thus, the aim of this study was to monitor the electrical conductivity of the composites during the water aging.

### 4. CONCLUSION

In this work, investigation was carried out on the effect of 2 wt. % of MWCNTs and 1–3 weight % of GNPs on the performance of nanocomposites to integrate advantages of the large surface area of GNPs and the high aspect ratio of MWCNTs. The findings are summarized as follows:
1. The geometries of MWCNTs and GNPs aided the dispersion by preventing the restacking of GNPs during wetting of glass fiber with epoxy blend and established a continuous conductive three-dimensional network. This increases the interface area between reinforcing glass fiber, fillers, and matrix, facilitating the transfer of stress, tension, and electrons across the interface.

2. The findings elucidate a marked synergistic effect between MWCNTs and GNPs, which may provide a guideline for the design and fabrication of GFRP epoxy composites using pultrusion using low fixed volume fractions of MWCNTs.
3. The result obtained confirms the synergy between GNPs and MWCNTs as both are carbon based and have supplementary geometries aiding GNC to be resistant to water aging. Thus, the study of water aging through monitoring the electrical conductivity is achieved.

**5. ACKNOWLEDGMENTS**

The authors would like to thank V.T.U Research Center, Department of EEE, S.I.T., Tumakuru, for providing financial assistance and an opportunity to carry out a research project on "Aging analysis of Nanocomposite core for High Voltage Transmission Line." The authors thank Electrical and Electronics Engineering, Physics and Chemistry Departments of Siddaganga Institute of Technology, Tumakuru, for encouraging and providing all the research facilities.

**6. REFERENCES**

1. M. R. Zakaria, M. H. Abdul Kudus, H. M. Akil, M. Z. M. Thirmizir, (2017) Comparative study of graphene nanoparticle and multiwall carbon nanotube filled epoxy nanocomposites based on mechanical, thermal and dielectric properties, *Composites Part B: Engineering*, **119:** 57-66.
2. S. Gantayat, D. Rout, S. K. Swain, (2018) Carbon nanomaterial reinforced epoxy composites: A review, *Polymer-Plastics Technology and Engineering*, **57:** 1-16.
3. A. Moosa, A. Ramazani, A. A. Moosa, M. N. Ibrahim, (2016) Mechanical and electrical properties of graphene nanoplates and carbon- nanotubes hybrid epoxy nanocomposites, *American Journal of Materials Science*, **6:** 157-165.
4. Z. Anwar, A. Kausar, I. Rafique, B. Muhammad, (2016) Advances in epoxy/graphene nanoplatelet composite with enhanced physical properties: A review, *Polymer-Plastics Technology and Engineering*, **55:** 643-662.
5. A. Al-Sabagh, E. Taha, U. Kandil, A. Awadallah, G. Abdelnaser, M. Nasr, M. R. Taha, (2017) Monitoring moisture damage propagation in GFRP composites using carbon nanoparticles, *Polymers* (*Basel*), **9:** 1-20.
6. J. R. Correia, M. Petrů, (2010) Effects of hygrothermal ageing on the mechanical properties of glass-fibre-reinforced polymer pultruded profiles, *Structural Engineering International*, **20:** 370-378.
7. G. Carra, V. Carvelli, (2014) Ageing of pultruded glass fibre reinforced polymer composites exposed to combined environmental agents, *Composite Structures*, **108:** 1019-1026.
8. B. Alemour, M. H. Yaacob, H. N. Lim, M. R. Hassan, (2018) Review of electrical properties of graphene conductive composites, *International Journal of Nanoelectronics and Materials*, **11:** 371-398.
9. I. Kranauskaitė, J. Macutkevic, A. Borisova, A. Martone, M. Zarrelli, A. Selskis, A. N. Aniskevich, J. Banys, (2017) Enhancing electrical conductivity of multiwalled carbon nanotube/epoxy composites by graphene nanoplatelets, *Lithuanian Journal of Physics*, **57:** 232-242.
10. D. E. Mouzakis, H. Zoga, C. Galiotis, (2008) Accelerated environmental ageing study of polyester/glass fiber reinforced composites (GFRPCs), *Composites Part B: Engineering*, **39:** 467-475.
11. D. Stengel, R. Bardl, C. Kuhnel, S. Grosmannn, W. Kiewitt, (2017) Accelerated electrical and mechanical ageing tests of high temperature low sag (HTLS) conductors. In: **12**$^{th}$ *International Conference on Live Maintenance*, *No.* 612748, p1-6.
12. G. M. Odegard, A. Bandyopadhyay, (2011) Physical ageing of epoxy polymers and their composites, *Journal of Polymer Science, Part B: Polymer Physics*, **49:** 1695-1716.
13. S. Sugiman, S. Salman, (2019) Hygrothermal effects on tensile and fracture properties of epoxy filled with inorganic fillers having different reactivity to water, *Journal of Adhesion Science and Technology*, **33:** 691-714.
14. C. Ramírez, M. Rico, A. Torres, L. Barral, J. López, B. Montero, (2008) Epoxy/POSS organic-inorganic hybrids: ATR-FTIR and DSC studies, *European Polymer Journal*, **44:** 3035-3045.